\documentclass[aps,pra,twocolumn,superscriptaddress,showpacs]{revtex4}
\usepackage{amsmath,amssymb,graphicx,times}
\usepackage[hypertex,colorlinks=true,bookmarks=true,citecolor=blue,linkcolor=blue,urlcolor=blue,breaklinks=true]{hyperref}

\newcommand{\PR}[4]{Phys. Rev. #1 {\bf #2}, 
\href{http://dx.doi.org/10.1103/PhysRev#1.#2.#3}{#3} (#4)}
\newcommand{\PRRC}[4]{Phys. Rev. #1 {\bf #2}, 
\href{http://dx.doi.org/10.1103/PhysRev#1.#2.#3}{#3(R)} (#4)}
\newcommand{\PRL}[3]{Phys. Rev. Lett. {\bf #1}, 
\href{http://dx.doi.org/10.1103/PhysRevLett.#1.#2}{#2} (#3)}

\newcommand{\JPSJ}[3]{J. Phys. Soc. Jpn. {\bf #1}, 
\href{http://dx.doi.org/10.1143/JPSJ.#1.#2}{#2} (#3)}
\newcommand{\bol}[1]{\boldsymbol #1}

\begin{document}
\title{Symmetry-protected topological order in magnetization plateau 
states of quantum spin chains}
\author{Shintaro Takayoshi}
\affiliation{Computational Materials Science Unit, National Institute for Materials Science, 
Namiki 1-1, Tsukuba, Ibaraki 305-0044, Japan}
\author{Keisuke Totsuka}
\affiliation{Yukawa Institute for Theoretical Physics, Kyoto University, 
Kitashirakawa Oiwake-Cho, Kyoto 606-8502, Japan}
\author{Akihiro Tanaka}
\affiliation{Computational Materials Science Unit, National Institute for Materials Science, 
Namiki 1-1, Tsukuba, Ibaraki 305-0044, Japan}

\date{\today}

\begin{abstract}

A symmetry-protected topologically ordered phase is 
a short-range entangled state, for which some imposed symmetry 
prohibits the adiabatic deformation into a trivial state 
which lacks entanglement. 
In this paper we argue that magnetization plateau states 
of one-dimensional antiferromagnets which satisfy the 
conditions $S-m\in$ odd integer, 
where $S$ is the spin quantum number and $m$ the 
magnetization per site, 
can be identified as symmetry-protected topological states 
if an inversion symmetry about the link center is present. 
This assertion is reached by mapping the 
antiferromagnet into a nonlinear sigma model type effective field theory 
containing a novel Berry phase term (a total derivative term) 
with a coefficient proportional to the quantity $S-m$, 
and then analyzing the topological structure of the ground state wave functional 
which is inherited from the latter term. A boson-vortex duality transformation is 
employed to examine the topological stability of the ground state 
in the absence/presence of a perturbation violating link-center inversion symmetry. 
Our prediction based on field theories is verified by means of a numerical study 
of the entanglement spectra of actual spin chains, 
which we find to exhibit twofold degeneracies 
when the aforementioned condition is met. 
We complete this study with a rigorous analysis using matrix product states.

\end{abstract}


\maketitle

\section{Introduction}

Classification of phases is an important and fundamental problem 
in statistical physics. 
In Ginzburg-Landau theories, 
different phases are distinguished by local order parameters. 
It has come to be realized in the past few decades, 
however, that there are phases 
which cannot be characterized by local order parameters but 
are nevertheless nontrivial.
Dubbed topologically ordered phases, they have in recent years become 
an intensively studied subject in condensed matter physics~\cite{Wen04}. 
There are largely two known types 
of topological orders. 
One is characterized by states with long-range 
entanglement that sustain anyonic excitations. 
The other arises in states with short-range entanglement 
and become robust as a phase 
when some specific symmetries are imposed; 
the imposed symmetry condition forbids 
perturbative terms, whose incorporation 
would otherwise smoothly deform the state  
into a direct product (i.e., topologically trivial) state.  
States characterized by the latter type of order are said to belong to 
a symmetry-protected topological (SPT) phase. 

\begin{figure}[t]
\includegraphics[width=0.4\textwidth]{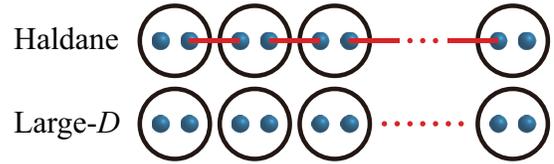}
\caption{(Color online) Schematic pictures of 
$S=1$ Haldane (VBS) and large-$D$ phases. 
The links connecting spin-1/2's (balls) represent singlet bonds.}
\label{fig:HaldaneVBS}
\end{figure}

A typical example of an SPT phase is 
the Haldane state~\cite{Haldane83,Affleck87} of $S=1$ quantum magnets, 
which is conveniently 
characterized as a valence bond solid (VBS) [Fig.~\ref{fig:HaldaneVBS}]. 
Haldane conjectured that Heisenberg antiferromagnets composed of  
integer spin have a nonzero excitation gap, while those with half-integer spin are gapless. 
Decades after this prediction was verified through many numerical calculations 
and experiments~\cite{Nightingale86,White93,Polizzi98,Katsumata89}, 
recent studies motivated by the quest for new topological phases of matter 
have made it apparent that 
even within the gapped ground states for integer spin systems, 
there is a qualitative difference between the $S=$ odd and even cases. 
For example, in the $S=1$ case, it was shown that 
a phase transition must always intervene between the Haldane phase and 
a topologically trivial phase (a typical example of the latter 
being the large-$D$ phase [Fig.~\ref{fig:HaldaneVBS}], 
 which can be expressed in the $S^{z}$ basis as $|\ldots0000\ldots\rangle$)    
provided one of the following three symmetries is imposed onto 
the system~\cite{Pollmann10}: 
(i) the dihedral group of $\pi$ rotations about the $x,y,z$ axes, 
(ii) time-reversal, and (iii) link-center inversion. 
For $S=2$ chains, in contrast, 
it is possible to connect the Haldane and large-$D$ phases  
adiabatically~\cite{Pollmann10}. 
Phases in one-dimensional gapped spin chains can thus be enriched 
by symmetry protection~\cite{Chen11} and their 
classification using group cohomology has been put forth~\cite{Chen12}. 
SPT phases in higher dimensions have also been 
proposed~\cite{Levin12,Lu12,Senthil13,Vishwanath13,Ye13}.

Gapped phases are also observed in antiferromagnets 
under an externally applied magnetic field. 
These are the magnetization plateaus,  
i.e., the regions within the  
magnetization curve 
where the magnetization remains unchanged 
with increasing field strength.  
The Oshikawa-Yamanaka-Affleck (OYA) theory generalizes the 
celebrated Lieb-Schultz-Mattis theorem~\cite{Lieb61} and  
summarizes the necessary condition for the appearance of a plateau 
in the form $r(S-m)\in\mathbb{Z}$, where 
$r$ is the number of sites in one unit cell and 
$S$ is the spin quantum number and $m$ the magnetization per site~\cite{Oshikawa97,Totsuka97}. 
The stability of this gapped plateau phase has also 
be explained in terms of an effective field theory where  
Berry phase terms play a crucial role~\cite{Tanaka09}. 
Strong analogies between the Haldane gap and  
magnetization plateau states have been noted early on~\cite{Oshikawa97}, 
and a VBS picture similar to Fig.~\ref{fig:HaldaneVBS} 
can be employed to depict the latter (see, e.g., Fig.~\ref{fig:PlateauVBS}). 
In view of the current interest in physical realizations 
of SPT states, it would thus seem important to search for magnetization 
plateau states which can be characterized as an SPT phase. This is the 
principal purpose of the present work. 
We note that there are several previous studies on 
magnetization plateaus which 
have also been conducted in the light of topological phases 
(though the explicit connection with SPT phases was not addressed): 
for instance, the Chern number and nontrivial edge excitations associated with 
plateau state  in periodically modulated chains was discussed in Ref.~\cite{Hu14}, 
while the entanglement spectra of plateau states which occur in  
ferro-ferro-antiferromagnetic (FFAF) chains was investigated in Ref.~\cite{Takayoshi14}. 

An additional motivation comes from 
the correspondence~\cite{Sachdev11,Tanaka09} between 
antiferromagnets subjected to a magnetic field and 
bosons with a tunable chemical potential (which can be casted into 
a Bose-Hubbard model~\cite{Fisher89}). SPT states 
which appear in the phase diagram of the Bose-Hubbard model have been 
proposed~\cite{DallaTorre06,Batrouni13}, and new developments in cold-atom 
experiments enable one to directly measure string orders~\cite{Endres11} 
which, in principle, detect the subtle topological orders which characterize 
such states. It is our hope that the present study will contribute some new insights 
to this closely related problem. 

This paper is organized as follows. 
We begin by developing in Sec.~\ref{sec:EffectiveFieldTheory}
an effective field theory for magnetization plateau states. 
In particular, we elicit a spin Berry phase term which will  
be central to the discussion that follows 
in Sec.~\ref{sec:TemporalEdge}. 
This term has a surface contribution which was not considered 
in the earlier field theoretical work 
described in Ref.~\cite{Tanaka09}. 
We then go on to investigate, 
following the path integral formalism of Ref.~\cite{Xu13}, 
the topological structure 
of the ground state wave functional of our spin system.  
Here we will see that the surface Berry phase term will 
contribute a phase factor to the wave functional 
through the presence of a {\it temporal surface term}, i.e., a temporal analog of the 
boundary Berry phase~\cite{Ng94} which represents the fractional spins 
which appear at the end of open spin chains. 
We show that this factor will 
govern whether or not the ground state belongs to an SPT phase. 
Our finding is that a plateau state can be in an SPT phase if $S-m$ is an odd integer. 
We also make clear that this SPT phase is protected 
by the link-center inversion (parity) symmetry $Z_{2}^{\rm P}$ 
by demonstrating that the protection of this phase 
is broken with the presence of staggered magnetic field. 
In Sec.~\ref{sec:Numerics}, we verify affirmatively this prediction by 
presenting numerical calculations for model spin systems 
giving rise to plateaus satisfying $S-m\in$ odd and $S-m\in$ even. 
This is carried out by examining the degeneracy of the 
entanglement spectrum, which provides a direct fingerprint 
of SPT order. 
Finally, in Sec.~\ref{sec:MPS} we present 
a matrix product state (MPS) construction 
for the $m=1/2$ plateau in $S=3/2$ chains, which enables us to 
confirm in a rigorous manner that this state is indeed an SPT phase. 
Section~\ref{sec:Summary} is devoted to discussions and a summary. 
In the Appendix, 
the classification of SPT phases protected by 
${\rm U(1)}\rtimes Z_{2}^{\rm P}$ is discussed using the MPS representation. 

\section{Effective field theory of magnetization plateau phases: topological terms}
\label{sec:EffectiveFieldTheory}

\begin{figure}[t]
\includegraphics[width=0.2\textwidth]{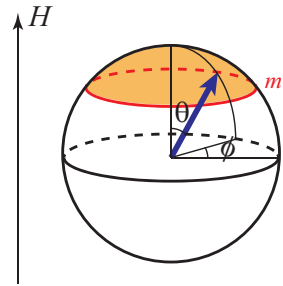}
\caption{(Color online) 
Schematic pictures of the spin Berry phase for magnetization plateau states 
in spherical coordinates.}
\label{fig:SphericalCoord}
\end{figure}

A field-theoretical description of the 
magnetization plateau state which emphasizes the 
role played by Berry phases was formulated in Ref.~\cite{Tanaka09}. 
In the following we refine this treatment in such a way 
that exposes the relation of this state to SPT phases. 
For simplicity, we hereon assume that a unit cell contains only one site, 
in which case the OYA condition reads $S-m\in\mathbb{Z}$. 
Following  Ref.~\cite{Tanaka09}, we shall start with the following 
Hamiltonian  which describes an 
antiferromagnetic spin chain in an external magnetic field, 
\begin{equation}
 {\cal H}=J\sum_{j}\bol{S}_{j}\cdot\bol{S}_{j+1}
   +D\sum_{j}(S_{j}^{z})^{2}-H\sum_{j}S_{j}^{z}.
 \label{eq:Hamil}
\end{equation}
We assume a canted spin configuration: 
\begin{equation}
 \bol{S}_{j}=S\bol{n}_{j}=
 \begin{pmatrix}
  \sqrt{S^{2}-m_{j}^{2}}\cos(kx)\\
  \sqrt{S^{2}-m_{j}^{2}}\sin(kx)\\
  m_{j}
\end{pmatrix}\quad
(x=ja),\label{eq:SpinCoord}
\end{equation}
where $a$ is a lattice constant and $k=\pi/a$. 
Spins are aligned in an antiparallel fashion within the $xy$ plane, 
while the $z$ component is uniform. 
As depicted in Fig.~\ref{fig:SphericalCoord}, 
we parametrize the corresponding unit vector $\bol{n}_{j}$ 
using spherical coordinates 
\begin{equation}
 \bol{n}_{j}(\tau)=
 \begin{pmatrix}
   (-1)^{j}\cos\phi_{j}(\tau)\sin\theta_{0}\\
   (-1)^{j}\sin\phi_{j}(\tau)\sin\theta_{0}\\
   \cos\theta_{0}
\end{pmatrix}.\nonumber
\end{equation}
The magnetization is $m=S\cos\theta_{0}$. 
For the classical solution, 
$\cos\theta_{0}=H/(2S(D+2J))$. 
We follow the treatment of Ref.~\cite{Tanaka09} in regard to the magnetization 
and hence $\theta_{0}$ for a given value of the magnetic field as fixed, 
taking into account the massive nature of the fluctuation of the latter quantity. 

The effective action for (\ref{eq:Hamil}) can be divided into 
the kinetic term ${\cal S}_{\rm kin}$ and the Berry phase 
contribution ${\cal S}_{\rm BP}^{\rm tot}$, 
\begin{equation}
 {\cal S}={\cal S}_{\rm kin}+{\cal S}_{\rm BP}^{\rm tot},\quad
 {\cal S}_{\rm kin}=\int d\tau{\cal H}.\label{eq:EffectiveAction}
\end{equation}
The continuum limit of the kinetic term is~\cite{Tanaka09} 
\begin{equation}
 {\cal S}_{\rm kin}\;\to\;
   \int d\tau dx\frac{\zeta}{2}
   \Big\{\frac{1}{v^{2}}(\partial_{\tau}\phi)^{2}
        +(\partial_{x}\phi)^{2}\Big\},\nonumber
\end{equation}
where
\begin{align}
 \zeta=&aJS^{2}
   \Big(1-\frac{H^{2}}{4S^{2}(D+2J)^{2}}\Big),\nonumber\\
 v=&Ja\sqrt{\frac{4S^{2}(D+2J)^{2}-H^{2}}{2J(D+2J)}}.\nonumber
\end{align}
The Berry phase part of the action (\ref{eq:EffectiveAction}) 
is the summation of the contribution from each site. 
For the sake of the following discussion, 
it proves convenient to introduce an auxiliary vector 
\begin{equation}
 \bol{N}_{j}(\tau)
\equiv
 \begin{pmatrix}
   \cos\phi_{j}(\tau)\sin\theta_{0}\\
   \sin\phi_{j}(\tau)\sin\theta_{0}\\
   \cos\theta_{0} 
 \end{pmatrix}.\label{eq:AuxVector}
\end{equation}
We note that the spin Berry phase term for $\bol{n}(\tau)$  
coincides with that for $\bol{N}_{j}(\tau)$. This follows since 
both spins precess in the same direction 
around the $z$ axis along the same constant latitude. 
The total Berry phase is 
${\cal S}_{\rm BP}^{\rm tot}=\sum_{j}{\cal S}_{\rm BP}[\bol{N}_{j}(\tau)]$, 
where 
\begin{align}
 {\cal S}_{\rm BP}[\bol{N}_{j}(\tau)]=&
   iS(1-\cos\theta_{0})
   \int d\tau\partial_{\tau}\phi_{j}\nonumber\\
   =&i(S-m)\int d\tau\partial_{\tau}\phi_{j}.
 \label{eq:SiteBerryPhase}
\end{align}
Applying the identity 
\begin{equation}
 {\cal S}_{\rm BP}[\bol{N}_{j}(\tau)]
   =2i(S-m)\int d\tau\partial_{\tau}\phi_{j}
    -{\cal S}_{\rm BP}[\bol{N}_{j}(\tau)]\nonumber
\end{equation}
only to terms associated with $j=$ even sites, 
we recast ${\cal S}^{\rm tot}_{\rm BP}$ as 
\begin{align}
 {\cal S}_{\rm BP}^{\rm tot}
   =&\sum_{j{\rm :odd}}{\cal S}_{\rm BP}[\bol{N}_{j}(\tau)]\nonumber\\
   &+\sum_{j{\rm :even}}\Big[2i(S-m)\int d\tau\partial_{\tau}\phi
     -{\cal S}_{\rm BP}[\bol{N}_{j}(\tau)]\Big]\nonumber\\
   =&\sum_{j}(-1)^{j}{\cal S}_{\rm BP}[\bol{N}_{j}(\tau)]+\sum_{j{\rm :even}}
     2i(S-m)\int d\tau\partial_{\tau}\phi_{j}.
 \label{eq:TotalBerryPhase}
\end{align}
In the continuum limit,
the second term of the last line of (\ref{eq:TotalBerryPhase}) reads 
\begin{equation}
 \sum_{j{\rm :even}}2i(S-m)\int d\tau\partial_{\tau}\phi_{j}\;\to\;
   i\int d\tau dx\frac{S-m}{a}\partial_{\tau}\phi.
 \label{eq:UniformBP}
\end{equation}
This is the Berry phase term 
that was derived in Ref.~\cite{Tanaka09}. 
There it was argued, by incorporating a     
a boson-vortex duality transformation, that   
for the case $S-m\notin\mathbb{Z}$, this term  
has a nontrivial effect and 
will generally lead to a gapless theory by prohibiting  
vortex condensation.   
Meanwhile, for $S-m\in\mathbb{Z}$, 
it was found that this term does not affect the low-energy physics, 
allowing for vortex proliferation and hence the formation 
of a gapped (i.e., the magnetization plateau) state. 
As we are focused on the latter situation, this term can be 
discarded. 

\begin{figure}[t]
\includegraphics[width=0.3\textwidth]{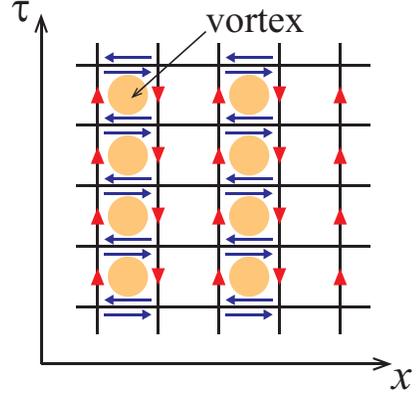}
\caption{(Color online) 
Following Ref.~\cite{Sachdev02}, 
we take a summation over spin Berry phases 
with an alternating sign (vertical arrows), which amounts to 
counting the vorticity in every other row. 
The horizontal arrows cancel due to the contributions  
from adjacent plaquettes.}
\label{fig:BPandVortex}
\end{figure}

We now turn to the first term of 
the last line of (\ref{eq:TotalBerryPhase}), 
which was previously not considered.  
By placing the system on a space-time grid, 
the summation over spin Berry phases 
with an alternating sign 
can conveniently be converted into a net space-time vorticity~\cite{Sachdev02} 
as schematically shown in Fig.~\ref{fig:BPandVortex}. We therefore have 
\begin{align}
 {\cal S}_{\rm BP}^{\rm tot}
   =&\sum_{j}(-1)^{j}{\cal S}_{\rm BP}[\bol{N}_{j}(\tau)]\nonumber\\
   =&i\sum_{j}(-1)^{j}(S-m)\int d\tau\partial_{\tau}\phi_{j}\nonumber\\
   =&i2\pi(S-m)\sum_{\substack{\rm odd\\\rm column}}
     ({\rm spacetime\;vorticity\;of}\;\phi).\nonumber
\end{align}
Taking the continuum limit, we obtain 
\begin{align}
 {\cal S}_{\rm BP}^{\rm tot}
   \to& i\frac{S-m}{2}\int d\tau dx
     (\partial_{\tau}\partial_{x}-\partial_{x}\partial_{\tau})
     \phi(\tau,x)\nonumber\\
\equiv&i\pi(S-m)Q_{\rm v},
 \label{eq:TotalBPSphereCoord}
\end{align}
where the factor 1/2 in the first line can be understood by 
observing that we are to add up the total 
vorticity in every other row, and $Q_{\rm v}$ is the 
net vorticity throughout the entire space-time.      

In order to gain an understanding on the physics which is represented by the 
action (\ref{eq:TotalBPSphereCoord}), 
it is insightful to compare this topological term with that which 
appears in the effective theory of a planar antiferromagnetic chain 
(without an applied magnetic field). It is well known that 
the O(3) nonlinear $\sigma$ (NL$\sigma$) model 
with a topological $\theta$ term 
captures the low-energy property of an antiferromagnetic spin chain. 
One way to incorporate the planar nature of the order parameter into this 
action while preserving the global topological properties of the theory, 
is to employ the CP$^{1}$ representation~\cite{Auerbach}, which re-expresses the 
planar unit vector 
\begin{equation}
 \bol{N}_{\rm pl}(\tau,x)\equiv
 \begin{pmatrix}
   \cos\phi(\tau,x)\\
   \sin\phi(\tau,x)\\
   0
 \end{pmatrix}\nonumber
\end{equation}
in terms of a spinor field $\bol{z}$ through the relation 
$N_{\rm pl}^{a}=\bol{z}^{\dagger}\sigma^{a}\bol{z}$ 
($\sigma^{a}$: Pauli matrices). 
Choosing the gauge 
\begin{equation}
 \bol{z}\equiv
 \begin{pmatrix}
   1/\sqrt{2}\\ e^{i\phi(\tau,x)}/\sqrt{2}
 \end{pmatrix},\nonumber
\end{equation} 
we find that the CP$^{1}$ gauge field is 
\begin{equation}
 a_{\mu}\equiv -i\bol{z}^{\dagger}\partial_{\mu}\bol{z}
   =\partial_{\mu}\phi/2 \;(\mu=\tau,x).\nonumber
\end{equation} 
Finally, plugging this into the CP$^{1}$ representation for the 
$\theta$ term~\cite{Auerbach}, 
\begin{equation}
 {\cal S}_{\theta}=i\frac{\theta}{2\pi}\int d\tau dx
   (\partial_{\tau}a_{x}-\partial_{x}a_{\tau}),\nonumber
\end{equation}
where the vacuum angle $\theta$ (not to be confused with the 
spherical coordinate) is $\theta=2\pi S$, we arrive at 
\begin{equation}
 {\cal S}_{\theta}^{\rm pl}=i\pi SQ_{\rm v}, 
   \label{eq:PlanarThetaTerm}
\end{equation}
which is consistent with the lattice-based results of Ref.~\cite{Sachdev02}. 
Also, as explained later in this section, it is possible to start from 
Eq.~(\ref{eq:PlanarThetaTerm}) to derive a dual vortex field theory 
which, when treated with due care, 
correctly discriminates the behavior of integer- and half-integer-$S$ systems 
in agreement with the Haldane conjecture. This fact lends credibility 
to the use of this particular expression in the continuum theory. 

Comparing Eqs. (\ref{eq:TotalBPSphereCoord}) and (\ref{eq:PlanarThetaTerm}), 
and noting in addition that the kinetic term of our action bears the form 
of an O(3) NL$\sigma$ model in the XY limit, 
we find that the two effective theories are identical in form. 
In particular, Eq.~(\ref{eq:TotalBPSphereCoord}) corresponds exactly 
to the $\theta$ term ${\cal S}_{\theta}^{\rm pl}$ with an effective vacuum angle of 
\begin{equation}
 \theta_{\rm eff}=2\pi(S-m).\nonumber
\end{equation} 
This coincidence is quite natural when we recall the VBS-construction 
of a magnetization plateau state, such as  depicted in Fig.~\ref{fig:PlateauVBS}. 
At each site, the dynamics of the polarized portion of the spin moment (of magnitude $m$) 
is pinned down by the magnetic field, while the subsystem consisting of the 
remaining ``active'' moment of magnitude $S-m$ 
form a VBS-like state. The low-energy physics of this state is therefore 
essentially that of (the planar limit of) a Haldane gap state of a 
spin $S-m$ antiferromagnet (recall that we are confining our attention 
to the case where $S-m\in\mathbb{Z}$). It is also worthwhile  
to note that the Berry phase action, Eq.~(\ref{eq:TotalBPSphereCoord}), is a 
total derivative term, which will have important consequences in the following 
section. 

In the remainder of this section, 
we discuss how the application of a 
duality transformation on the effective action obtained above: 
\begin{align}
 {\cal S}=&{\cal S}_{\rm kin}
   +{\cal S}_{\rm BP}^{\rm tot},\nonumber\\
 {\cal S}_{\rm kin}=&\int d\tau dx\frac{\zeta}{2}
   \Big\{\frac{1}{v^{2}}(\partial_{\tau}\phi)^{2}
        +(\partial_{x}\phi)^{2}\Big\},
 \label{eq:PlateauEffAction}\\
 {\cal S}_{\rm BP}^{\rm tot}=&
   i(S-m)\int d\tau dx
   (\partial_{\tau}a_{x}-\partial_{x}a_{\tau}),\nonumber
\end{align}
will enable us to seek additional insight 
from the viewpoint of vortex condensates. 
Since (i) the action (\ref{eq:PlateauEffAction}) 
is basically a quantum XY model, and 
(ii) vortex proliferation is allowed in the magnetization plateau phase, 
it is not difficult to guess that the dual vortex-field theory will come out 
as some variant of the quantum sine-Gordon action, a fact which will be 
verified shortly.  
Though the technicalities involved in carrying out the ``dual-izing'' 
procedure were described in some details in Ref.~\cite{Tanaka09}, we briefly 
sketch the main steps as there are differences stemming from use of a 
different topological term. 
For the sake of simplicity, the mapping will be performed in the continuum limit. 
The corresponding procedures can, however, be carried out on the lattice as well, 
which can easily be seen to lead to identical results. 

We start then, with a slight rewriting of the effective Lagrangian density: 
\begin{equation}
 {\cal L}=\frac{1}{2g}(\partial_{\mu}\phi)^{2}
   +i\pi(S-m)\rho_{\rm v},\nonumber
\end{equation}
where $\rho_{\rm v}\equiv
(\partial_{\tau}\partial_{x}-\partial_{x}\partial_{\tau})\phi/(2\pi)$ 
is the density of spacetime vortices.
Here we have set $v=1$ and $g=1/\zeta$ for notational simplicity. 
A Hubbard-Stratonovich transformation recasts 
the kinetic term as 
$(\partial_{\mu}\phi)^{2}/(2g)\to(g/2)J_{\mu}^{2}+iJ_{\mu}\partial_{\mu}\phi$. 
After decomposing $\phi$ into 
a vortex-free portion $\phi_{\rm r}$ 
and a portion with vorticity $\phi_{\rm v}$, i.e., 
$(\partial_{\tau}\partial_{x}-\partial_{x}\partial_{\tau})\phi_{\rm r}=0$ 
and
$(\partial_{\tau}\partial_{x}-\partial_{x}\partial_{\tau})\phi_{\rm v}\neq 0$, 
the integration over $\phi_{\rm r}$ 
yields a delta function contribution $\propto\delta(\partial_{\mu}J_{\mu})$. 
The constraint $\partial_{\mu}J_{\mu}=0$ is formally 
solved by introducing a new vortex-free scalar field $\varphi$ 
and putting $J_{\mu}\equiv\epsilon_{\mu\nu}\partial_{\nu}\varphi/(2\pi)$, 
which leads to
\begin{equation}
 {\cal L}=\frac{g}{8\pi^{2}}(\partial_{\mu}\varphi)^{2}
   +i\big[\pi(S-m)-\varphi\big]\rho_{\rm v}.\nonumber
\end{equation}
Integrating out the dual field $\varphi$, we obtain
\begin{equation}
 {\cal L}_{\rm dual}=\frac{2\pi^{2}}{g}\rho_{\rm v}\frac{1}{-\partial^{2}}
   \rho_{\rm v}+i\pi(S-m)\rho_{\rm v}.\nonumber
\end{equation} 
This is the dual action for vortices written 
at the first quantization level, 
describing the intervortex logarithmic interaction (first term) 
and the vortex Berry phase (second term). 

Alternatively one can proceed to derive a vortex field theory, 
i.e., a dual theory at the second quantization level. 
For this purpose, we submit the system to a 
standard fugacity expansion and restrict the vorticity to $\pm 1$. 
Denoting the fugacity as $z=e^{-\mu}$, where $\mu$ is the 
creation energy of a vortex,  the grand canonical partition function of the vortex gas 
becomes 
\begin{widetext}
\begin{align}
 Z=&\int{\cal D}\varphi
   e^{-\int d^{2}\bol{x}(g/(8\pi^{2}))(\partial_{\mu}\varphi)^{2}}
   \sum_{N_{+}=0}^{\infty}\sum_{N_{-}=0}^{\infty}
   \frac{(ze^{ i\pi(S-m)})^{N_{+}}}{N_{+}!}
   \frac{(ze^{-i\pi(S-m)})^{N_{-}}}{N_{-}!}
   \Big(\int d^{2}\bol{r}_{i_{+}}e^{-i\varphi(\bol{r}_{i_{+}})}\Big)^{N_{+}}
   \Big(\int d^{2}\bol{r}_{i_{-}}e^{ i\varphi(\bol{r}_{i_{-}})}\Big)^{N_{-}}
 \nonumber\\
 =&\int{\cal D}\varphi
   \exp\Big[-\int d^{2}\bol{x}\Big\{
     \frac{g}{8\pi^{2}}(\partial_{\mu}\varphi)^{2}
     +2z\cos\big[\pi(S-m)-\varphi\big]\Big\}\Big].\nonumber
\label{eq:partition function}
\end{align}
\end{widetext}
Taking into account the condition $S-m\in\mathbb{Z}$, 
the final form of the vortex field theory reads 
\begin{equation}
 {\cal L}_{\rm dual}=\frac{g}{8\pi^{2}}(\partial_{\mu}\varphi)^{2}
   +2z\cos\big[\pi(S-m)\big]\cos\varphi.
 \label{eq:SineGordon}
\end{equation}
Having started with a new 
Berry phase term which only applies to the plateau state, 
our vortex field theory   
differs from the one found in Ref.~\cite{Tanaka09}.
The Berry phase for each space-time vortex event 
is reflected in the coefficient $\cos\big[\pi(S-m)\big]$ 
of the cosine term.  
This cosine term has scaling dimension $\pi/g$, 
and becomes relevant when $g>\pi/2$, 
in which regime the field $\varphi$ ($\phi$) is ordered (disordered),  
due to vortex proliferation.
Because of the sign dependence of the cosine term  
in (\ref{eq:SineGordon}) on the parity of $S-m$, 
a different value of $\varphi$ is favored   
depending on whether $S-m$ is odd or even. 
While this suggests that the two cases describe 
two distinct phases~\cite{Fuji14}, 
we shall postpone their characterization 
as SPT or trivial phases until 
Sec.~\ref{sec:TemporalEdge}, 
where a fuller picture will emerge by relating 
the behavior of the dual theory (\ref{eq:SineGordon}) 
to the topological 
structure of the ground state wave functional. 

One thing that follows immediately from the above action is that
a phase-soliton of height $\pi$  
must exists at a junction of $S-m=$ even and odd systems. 
We can surmise that this soliton carries a spin-1/2, 
which would correspond to a boundary spin. 
This is conveniently seen by refermionizing the bosonic field $\varphi$ to 
right and left moving fermions $R,L$.  
For simplicity, let us suppose that the Luttinger liquid parameter $g$ is tuned such that 
the first term in (\ref{eq:SineGordon}) corresponds to the free Dirac fermion 
(i.e., $g=\pi$). 
The Lagrangian (\ref{eq:SineGordon}) then 
is equivalent to the following massive Dirac fermion 
\begin{equation}
 {\cal L}_{\rm fer}=\bar{\psi}\gamma^{\mu}\partial_{\mu}\psi
   +M\bar{\psi}\psi,\nonumber
\end{equation}
where $\psi^{\rm T}=(R,L)$, $\gamma_{0}=\sigma_{1}$, $\gamma_{1}=-\sigma_{2}$, 
$\gamma_{5}=\sigma_{3}=i\gamma_{0}\gamma_{1}$, 
$\bar{\psi}=\psi^{\dagger}\gamma_{0}$, and 
$M=16\pi z\cos\big(\pi(S-m)\big)$
($\sigma_{1,2,3}$ are Pauli matrices).
Through the relations: 
\begin{align}
 i(\partial_{\tau}-i\partial_{x})\varphi/(4\pi)
   \leftrightarrow& :R^{\dagger}R:\nonumber\\
 -i(\partial_{\tau}+i\partial_{x})\varphi/(4\pi)
   \leftrightarrow& :L^{\dagger}L:,\nonumber
\end{align}
the charge density of fermions is represented as 
\begin{equation}
 \rho\equiv:R^{\dagger}R+L^{\dagger}L:=\partial_{x}\varphi/(2\pi).
 \nonumber
\end{equation}
For $S-m=$ even (odd) systems, 
the fermion mass $M$ is positive (negative) 
and $\varphi$ is pinned at $\pi$ (0). 
If $S-m=$ even (odd) in $x>0$ ($x\leq0$), 
the fermion charge is 
\begin{equation}
 \int_{-\infty}^{\infty}dx\rho
   =\int_{-\infty}^{\infty}dx\frac{\partial_{x}\varphi}{2\pi}
   =\frac{1}{2}.\nonumber
\end{equation}
This implies that a soliton with fractionalized charge 
exists at the boundary between $S-m=$ odd and even systems. 
This soliton corresponds to the boundary spin-1/2 stated above. 
The action (\ref{eq:SineGordon}) also tells us that, even for the same sign of $z$, we may have 
different types of plateaus depending on the parity of $S-m$.  
We will see examples of this in Sec.~\ref{sec:Numerics}. 

Finally we mention that a quantum XY model with the Berry phase term 
of Eq.~(\ref{eq:PlanarThetaTerm}), which describes 
planar antiferromagnetic chains in the 
absence of external fields, can be 
submitted to the same duality procedures of the 
proceeding paragraphs. 
The final vortex field theory is identical in form to  
the sine-Gordon action (\ref{eq:SineGordon}), 
but with $S-m$ replaced by $S$. 
(Here a slight subtlety must be accounted for to reach this form. 
Namely, we need to assume for this purpose that the system is 
an easy-plane antiferromagnet, and the spin would prefer to 
escape into the out-of-plane direction at the vortex core, which would be 
less costly than creating a singular core. 
It is important to notice at this point that for 
a given winding sense of a vortex configuration, 
there are two possible (up and down)  directions in which the spin at the core 
can point. Taking all four topological (meron) configurations into account, 
we arrive at the dual theory mentioned above~\cite{Affleck89}.) 
If $S$ is a half-odd integer, the coefficient $\cos(\pi S)$ vanishes identically, yielding 
a massless theory, while the cosine term can generate a mass gap for integer $S$.    
The consistency of this result with the Haldane conjecture provides 
a useful check on the validity of the type of field theory 
that we have discussed in this section.

\section{Temporal surface terms and the SPT order in magnetization plateau states}
\label{sec:TemporalEdge}

Having arrived at our effective action (\ref{eq:PlateauEffAction}), 
we are in a position to discuss, 
along the lines of Ref.~\cite{Xu13}, 
the possible emergence of SPT order 
in a magnetization plateau state. 
The basic strategy is to express the ground state wave functional 
using a Feynman path integral representation, 
and thereby isolate the phase factor which comes 
from a temporal surface contribution generated 
by the Berry phase term. 
(Imposing a periodic boundary condition along the spatial direction, as we will do below, 
implies that the action does not contain the more familiar 
spatial surface contributions.) 
One finds that this {\it temporal surface term} 
encodes into the wave functional the 
information necessary to discriminate between states 
with and without the SPT order. 

Here we will take the strong coupling limit $\zeta\to 0$, 
and only consider the topological part of the action, 
although this is not an absolute necessity 
and will not effect the conclusions.  
Focusing primarily on bulk properties, we will employ 
spatial periodic boundary conditions. 
The state vector for the ground state can be expanded with respect to the  
basis $|\bol{N}_{\rm pl}(x)\rangle$, which we use as a shorthand notation 
for the spin coherent state corresponding to the snapshot configuration 
$\{\bol{N}_{\rm pl}(x)\}$, 
\begin{equation}
 |\Psi_{\rm GS}\rangle=\sum_{\bol{N}_{\rm pl}(x)}
   \Psi[\bol{N}_{\rm pl}(x)]|\bol{N}_{\rm pl}(x)\rangle.
 \nonumber
\end{equation}
Each coefficient (wave functional) $\Psi[\bol{N}_{\rm pl}(x)]$ 
stands for the probability amplitude that the configuration $\{\bol{N}_{\rm pl}(x)\}$ 
occurs in the ground state, and can formally be expressed in path integral 
language as an evolution in imaginary time starting 
from some initial configuration:
\begin{equation}
 \Psi[\bol{N}_{\rm pl}(x)]\propto
   \int_{\bol{N}_{\rm pl,i}}^{\bol{N}_{\rm pl,f}=\bol{N}_{\rm pl}(x)}
   {\cal D}\bol{N}'_{\rm pl}(\tau,x)e^{-{\cal S}[\bol{N}'_{\rm pl}(\tau,x)]}.
 \label{eq:CoeffPathInt}
\end{equation}
Here, $\bol{N}_{\rm pl,i(f)}\equiv\bol{N}'_{\rm pl}(\tau_{\rm i(f)},x)$ 
represents spin configurations at initial (final) imaginary time $\tau_{\rm i(f)}$. 
Substituting the expression for the Berry phase term 
in (\ref{eq:PlateauEffAction}) into (\ref{eq:CoeffPathInt}), 
and taking into account the spatial periodic boundary condition yields 
\begin{align}
 \Psi[&\bol{N}_{\rm pl}(x)]\nonumber\\
   &\propto\int_{\bol{N}_{\rm pl,i}}^{\bol{N}_{\rm pl}(x)}
   {\cal D}\bol{N}'_{\rm pl}(\tau,x)
   e^{-i(S-m)\int dx[a_{x}(\tau_{\rm f})-a_{x}(\tau_{\rm i})]}.
 \nonumber
\end{align}
Localized at the two ends of the interval on which the imaginary-time integration is 
performed, the exponent in the right-hand side of the above equation may be viewed 
as the action contributed by temporal surface terms, as already mentioned. 
Since $a_{x}(\tau_{\rm i})$ is fixed by the initial condition and can be 
placed outside the path integral (a Feynman sum over initial configurations is to be 
performed afterwards), we need only concern ourselves with the term 
involving $a_{x}(\tau_{\rm f})$, and we are left with
\begin{equation}
 \Psi[\bol{N}_{\rm pl}(x)]\propto e^{-i(S-m)\pi W}, 
 \label{eq:GSwf}
\end{equation}
where $W\equiv(1/2\pi)\int dx\partial_{x}\phi\in\mathbb{Z}$ is 
the winding number which records the number of times  
the planar vector $\bol{N}_{\rm pl}(x)$ 
wraps around its circular target space as we follow its orientation 
along the spatial extent of the system.

For $S-m={\rm even}$, $e^{i(S-m)\pi W}\equiv1$ 
and the wave functional~(\ref{eq:GSwf}) reduces to that 
in the absence of the topological term. 
The case where $S-m=$ odd, meanwhile, yields the nontrivial 
factor $(-1)^{W}$.  
This suggests that the ground states break up into two sectors, i.e., 
that odd-($S-m$) plateaus are SPT states that are topologically distinct 
from the even-($S-m$) states, which we expect to be topologically trivial.  
The construction of SPT states by modifying signs of the trivial wave functional 
is close in spirit to the construction of the Ising-like SPT in two dimensions 
from a trivial paramagnetic state~\cite{Levin12}. 
To verify this assertion it remains to identify, as we will address in a moment, 
the symmetry (or symmetries) 
which can protect this distinction (i.e., prevent the addition of 
perturbations that will 
cause the wave functional to adiabatically interpolate between 
the above two sectors). 

We mention in passing that we have restricted our attention to the 
case where the unit cell consists of one site. 
The extension of our treatment to a system with $r$ sites per cell 
is straightforward. 
There the parity of 
\begin{equation}
 r(S-m)\in\mathbb{Z}\label{eq:SPTConditionMulti}
\end{equation}
(where $m$ still stands for the magnetization per site) will replace the 
role of $S-m$ of the present argument. 

\begin{figure}[t]
\includegraphics[width=0.45\textwidth]{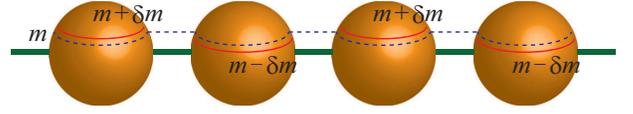}
\caption{(Color online) 
Application of a staggered magnetic field along the $z$ axis alters 
the local magnetization from $S-m$ to $S-m+(-)^{j}\delta m$, 
which clearly results in the 
breakdown of the structure~(\ref{eq:GSwf}). 
This perturbation can be prohibited by imposing 
a link-center inversion symmetry on the system.}
\label{fig:Perturbation}
\end{figure}

Let us now consider the effect of applying a staggered magnetic field 
along the $z$ axis [Fig.~\ref{fig:Perturbation}]. 
Repeating the derivation of the total Berry phase for this case, 
it is clear that this is the generic perturbation that 
directly affects the first (sign-alternating) 
term of (\ref{eq:TotalBerryPhase}), leading to the 
modification $S-m\rightarrow S-m-\delta m$, 
while leaving unchanged the second (uniform) term. 
The latter contribution can therefore be discarded 
as before, and we are left with 
\begin{equation}
 {\cal S}_{\rm BP}^{\rm tot}=i(S-m-\delta m)\int d\tau dx
   (\partial_{\tau}a_{x}-\partial_{x}a_{\tau}).\nonumber
\end{equation}
Accordingly, the wave functional formerly expressed 
by (\ref{eq:GSwf}) now takes on the form 
\begin{equation}
 \Psi[\bol{N}_{\rm pl}(x)]\propto
   e^{-i(S-m-\delta m)\pi W}.\nonumber
\end{equation}
By varying $\delta m$, it is now possible to continuously interpolate between the 
two aforementioned dependencies of the wave functional on the winding number $W$. 
 Additional information comes from 
revisiting the derivation of the vortex field theory (\ref{eq:SineGordon}); 
upon applying the staggered magnetic field, the action modifies to 
\begin{equation}
 {\cal L}_{\rm dual}
   =\frac{g}{2}(\partial_{\mu}\varphi)^{2}
    +2z\cos\big[\pi(S-m-\delta m)-\varphi\big].\nonumber
\end{equation}
It is clear from this action that the optimal value of the field $\varphi$ will change continuously 
as $\delta m$ is varied, without ever closing the energy gap. 
We therefore find that the 
application of a staggered magnetic field ruins the topological distinction 
between the even- and odd-($S-m$) cases.
Since this perturbation can be prohibited by requiring 
that the system respect inversion symmetry 
with respect to the center of a link, 
our observations strongly suggest that the  
odd-($S-m$) plateaus are SPT states protected by link-center inversion symmetry 
and are distinct from the even-($S-m$) plateaus, which are topologically trivial. 
We will arrive at the same conclusion both through 
a numerical study in Sec.~\ref{sec:Numerics}, and 
by analyzing an MPS representation for magnetization plateaus 
in Sec.~\ref{sec:MPS}.

\section{Numerical calculations}
\label{sec:Numerics}

In this section, we numerically provide an independent confirmation of 
the prediction by field theories that 
the parity of $S-m$ determines whether the system is in SPT phase or not. 
A simple way to distinguish SPT and trivial phases is 
to investigate the degeneracy of the entanglement spectrum~\cite{Pollmann10}. 
A numerical means suited for this purpose is  
the infinite time-evolving block decimation (iTEBD)~\cite{Vidal07}. 
This method utilizes the ability of MPS to approximate gapped states  
and enables one to simulate {\em infinite-size} systems 
by assuming a certain sort of spatial periodicity in the ground state. 
Through the imaginary-time evolution, the state approaches the ground state optimally 
approximated within the MPS representation with a fixed matrix dimension $\chi$. 
In general, the true ground state is better approximated with larger $\chi$. 

\begin{figure}[t]
\includegraphics[width=0.4\textwidth]{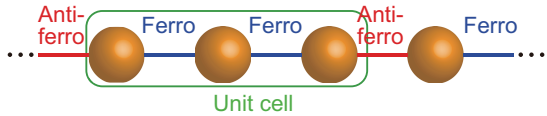}
\caption{(Color online) 
An illustration of an FFAF chain. Each circle represents a spin-$S$ object. 
Two successive ferromagnetic couplings and one antiferromagnetic 
coupling residing on the nearest neighbor links are repeated as depicted.}
\label{fig:FFAF}
\end{figure}

An ideal quantum spin model for studying SPT phases 
in plateaus would be a Heisenberg model 
with single-ion anisotropy $D$ [Eq.~(\ref{eq:Hamil})]. 
In fact, the existence of two different kinds of $m=1/2$ plateaus 
has been pointed out in this model with $S=3/2$~\cite{Kitazawa00}. 
It has been also argued there that the one appearing for small $D$ 
(i.e., $0.387\leq D/J \leq 0.943$) 
is characterized by the VBS-like state 
(where magnetization appears in the 
background of a VBS-like state) 
shown in Fig.~\ref{fig:PlateauVBS}(a).  
In addition, when $D$ further increases, a second-order Gaussian transition occurs and 
the system enters another plateau phase reminiscent of 
the large-$D$ phase, in which the unmagnetized background spin moments 
are quenched [Fig.~\ref{fig:PlateauVBS}(b)], and the short-range entanglement is absent. 

It is easy to understand this fact 
in terms of our effective action (\ref{eq:SineGordon}). 
Observing that the sign of the coefficient  
$z\cos\big[\pi(S-m)\big]$ of the cosine term in (\ref{eq:SineGordon}) 
determines whether the system in question is topological or not, one immediately sees that 
the transition between the VBS-like and large-$D$ plateaus studied in Ref.~\cite{Kitazawa00} 
should be characterized by $z=0$.  
This explains why the phase transition is of the Gaussian type. 
We note that this mechanism is essentially the same as that for 
the transition between the Haldane and large-$D$ phases   
in $S=1$ chains~\cite{Schulz86,Chen03}. 
Unfortunately, this plateau appears only in a tiny window of the applied magnetic field $H$ 
and it is difficult to approach this state by iTEBD 
simulations with fixed $H$. 

In order to circumvent 
the technical difficulties mentioned above, 
we study the following FFAF Heisenberg 
chain~\cite{Hida94} [see Fig.~\ref{fig:FFAF}]: 
\begin{align}
 {\cal H}_{\text{FFAF}}=&\sum_{j=1}^{L}
      (-J_{\text{F}}\bol{S}_{3j-2}\cdot\bol{S}_{3j-1}
       -J_{\rm  F}\bol{S}_{3j-1}\cdot\bol{S}_{3j  }\nonumber\\
&\qquad+J_{\rm AF}\bol{S}_{3j  }\cdot\bol{S}_{3j+1})
       -HS_{\rm tot}^{z},\nonumber
\end{align}
where $L$ is the number of unit cells and 
$S_{\rm tot}^{z}\equiv\sum_{j=1}^{3L}S_{j}^{z}$. 
Coupling constants $J_{\rm F}$ and $J_{\rm AF}$ 
are both positive. 
Each unit cell consists of three spins which are 
coupled through ferromagnetic bonds [Fig.~\ref{fig:FFAF}]. 
Due to the Hund rule coupling $-J_{\text{F}}$, 
we regard this unit cell as one spin-$3S$. 
Magnetization $m$ corresponds to 
$\langle S_{\rm tot}^{z}\rangle/L$. 
Thus, the situation is the same as considering 
plateaus with magnetization $\langle S_{\rm tot}^{z}\rangle/L$ 
in spin-$3S$ chains. 
We can give an equivalent description by substituting 
$r=3$ and $m=\langle S_{\rm tot}^{z}\rangle/(3L)$
into (\ref{eq:SPTConditionMulti}). 

\begin{figure}[t]
\includegraphics[width=0.3\textwidth]{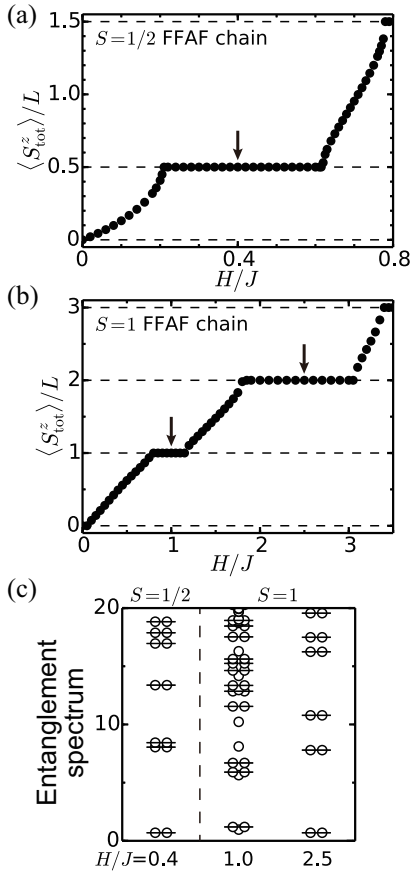}
\caption{Magnetization curves of (a) $S=1/2$ and (b) $S=1$ FFAF chains. 
(c) Entanglement spectra obtained by the bipartition at  
an antiferromagnetic bond under $H/J=0.4$ for $S=1/2$, 
$H/J=1.0$ and 2.5 for $S=1$ (shown by arrows in panels (a), (b)). 
Calculations are performed using iTEBD.}
\label{fig:MCurve_EntSpec}
\end{figure}

In particular, we performed calculations 
for $S=1/2$ and $S=1$ FFAF chains. 
We set the couplings at $J_{\rm F}=J_{\rm AF}=J$ for $S=1/2$ 
and $J_{\rm F}=J$, $J_{\rm AF}=2J$ for $S=1$. 
The magnetization curves of the $S=1/2$ and $S=1$ are 
shown in Figs.~\ref{fig:MCurve_EntSpec}(a) and 
\ref{fig:MCurve_EntSpec}(b). 
We employed the iTEBD method with MPS dimensions of  
$\chi=150$ and 100 for $S=1/2$ and $S=1$, respectively. 
Magnetization plateaus appear 
at $\langle S_{\rm tot}^{z}\rangle/L=1/2$ for $S=1/2$ 
and at $\langle S_{\rm tot}^{z}\rangle/L=1,2$ for $S=1$ satisfying the OYA condition 
$3S-\langle S_{\rm tot}^{z}\rangle/L\in\mathbb{Z}$. 

In order to check if the above plateaus 
are topologically nontrivial, we investigate 
the entanglement spectrum~\cite{Li08} which is known to give the fingerprints 
of topological phases. 
The entanglement spectrum of a quantum state $|\Psi\rangle$ 
is defined through the bipartition of the system into regions A and B. 
The state $|\Psi\rangle$ can be Schmidt-decomposed as 
a superposition of direct products 
$|\Psi\rangle=\sum_{\alpha}\lambda_{\alpha}
|\Psi_{\rm A}\rangle_{\alpha}\otimes|\Psi_{\rm B}\rangle_{\alpha}$ 
using the orthonormal basis sets $\{|\Psi_{\text{A}}\rangle_{\alpha}\}$ 
and $\{|\Psi_{\text{B}}\rangle_{\alpha}\}$ 
of the subsystems. 
The entanglement spectrum is defined by the logarithm 
$\{-\ln(\lambda_{\alpha}^{2})\}$ $(\alpha=1,\ldots,\chi)$ of 
the Schmidt eigenvalues $\lambda_{\alpha}$ 
normalized as $\sum_{\alpha}\lambda_{\alpha}^{2}=1$. 
As will be discussed in Sec.~\ref{sec:MPS}, if our system is in the SPT phase,  
the entanglement spectrum is twofold degenerate, 
in other words, all the values $\lambda_{\alpha}$ should appear in pairs. 
We emphasize that the bipartition of the system 
should be made at an antiferromagnetic bond 
since ferromagnetically coupled three spins 
are considered as one site. 
In Fig.~\ref{fig:MCurve_EntSpec}(c), we present the entanglement spectra 
obtained for the plateau states at magnetic fields $H/J=0.4$ for $S=1/2$ and 
$H/J=1.0$ and 2.5 for $S=1$ 
[shown by arrows in Figs.~\ref{fig:MCurve_EntSpec}(a) and 
\ref{fig:MCurve_EntSpec}(b)]. 
We can clearly see that entanglement spectra at $H/J=0.4$ 
($\langle S_{\rm tot}^{z}\rangle/L=1/2$) for $S=1/2$ 
and at $H/J=2.5$ ($\langle S_{\rm tot}^{z}\rangle/L=2$) 
for $S=1$ exhibit twofold degeneracy 
while the one at $H/J=1.0$ ($\langle S_{\rm tot}^{z}\rangle/L=1$) for $S=1$ does not. 
These results imply that the system is in the SPT (trivial) phase 
for $3S-\langle S_{\rm tot}^{z}\rangle/L=$ odd (even), thus confirming 
the prediction from quantum field theories discussed 
in Sec.~\ref{sec:TemporalEdge}. 

In closing this section, we briefly remark on the connection between 
the findings of this section with the field theoretical study of the 
earlier sections.  
In Sec.~\ref{sec:TemporalEdge}, we saw that the global structure of the 
ground state wave functional is determined by a temporal surface contribution 
coming from the topological term of the effective action. We can formally 
express the reduced density matrix 
$\rho_{\rm A}={\rm Tr}_{\rm B}|\Psi\rangle\langle\Psi|$
(where $|\Psi\rangle$ is the ground state and the trace operation 
is to be restricted to the region B),  
an object from which the entanglement spectrum can directly be extracted, 
along the same lines by incorporating a path integral 
representation~\cite{Nishioka09}. 
Once again the topological term 
will give rise to a surface contribution, this time along a segment 
of the imaginary time axis bounded by the spatial 
edge of region A. 
While we expect that this will play an essential role in determining the 
entanglement spectrum, we leave the details for future work.

\section{MPS representation of plateau phases}
\label{sec:MPS}

\begin{figure}[t]
\includegraphics[width=0.45\textwidth]{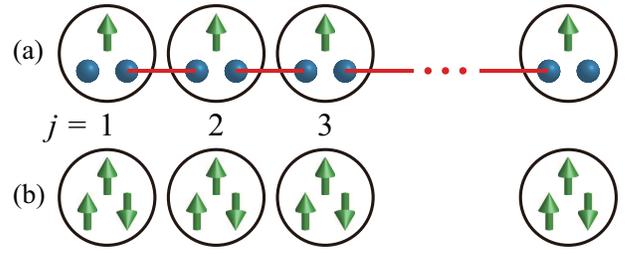}
\caption{(Color online) 
(a) The VBS picture of a model plateau state \eqref{eq:VBSPlateau} at $m=1/2$ 
in a $S=3/2$ spin chain. (b) Large-$D$ plateau.}
\label{fig:PlateauVBS}
\end{figure}

In this section, we present a simple model ground state (MPS) 
that exhibits the SPT properties 
at finite magnetic fields and show that the degenerate structure 
found above in the entanglement spectrum 
is indeed closely tied to the underlying topological properties. 
We consider the $m=1/2$ plateau in a $S=3/2$ chain for example. 
On top of the trivial product state
\begin{equation} 
 \bigotimes_{j} |S_{j}^{z}=1/2\rangle \; ,\nonumber
\end{equation}
we can think of an entangled state described by the VBS picture, 
schematically shown in Fig.~\ref{fig:PlateauVBS}.  
Using the auxiliary (Schwinger) bosons,   
this state is represented as 
\begin{equation}
 |\Psi\rangle=\prod_{j} a_{j}^{\dagger}(a_{j}^{\dagger}b_{j+1}^{\dagger}
   -b_{j}^{\dagger}a_{j+1}^{\dagger})\bigotimes_{j}|0\rangle_{j},
 \label{eq:VBSPlateau}
\end{equation}
where $a_{j}^{\dagger}$ and $b_{j}^{\dagger}$ are the bosonic creation operators 
of spin-1/2 up and down on the $j$th site, respectively.  
Note that there are exactly three bosons at each site which guarantee a local spin-3/2 at each site.  
This is the exact ground state of the following $S=3/2$ Hamiltonian~\cite{footnote1}: 
\begin{align}
 {\cal H}_{\rm VBS}=\sum_{j}\Big\{
    \bol{S}_{j}\cdot\bol{S}_{j+1}
   +&\frac{116}{243}(\bol{S}_{j}\cdot\bol{S}_{j+1})^{2}\nonumber\\
   +&\frac{ 16}{243}(\bol{S}_{j}\cdot\bol{S}_{j+1})^{3}
 \Big\}
\label{eq:ParentHamMagnetizedVBS}
\end{align}
with local magnetization 1/2.  

The MPS representation of the above state is given by 
\begin{equation}
 |\Psi\rangle=C\sum_{S_{j}^{z}=-3/2}^{3/2}
   \cdots A[S_{j-1}^{z}]A[S_{j}^{z}]A[S_{j+1}^{z}]\cdots
   \bigotimes_{j}|S_{j}^{z}\rangle,
 \label{eq:MPS}
\end{equation}
where
\begin{alignat}{2}
 &A[ 3/2]=\begin{pmatrix}0&0       \\-\sqrt{6}&0\end{pmatrix},&\quad
 &A[ 1/2]=\begin{pmatrix}\sqrt{2}&0\\0&-\sqrt{2}\end{pmatrix},\nonumber\\
 &A[-1/2]=\begin{pmatrix}0&\sqrt{2}\\0&0        \end{pmatrix},&\quad
 &A[-3/2]=\begin{pmatrix}0&0       \\0&0        \end{pmatrix} \nonumber
\end{alignat}
and $C$ is a normalization constant. 
The transfer matrix of the MPS~(\ref{eq:MPS}) is readily obtained as 
\begin{align}
 T_{(\alpha_{1},\alpha_{2}),(\beta_{1},\beta_{2})}\equiv &
   \sum_{S_{j}^{z}=-3/2}^{3/2}
     A_{\alpha_{1},\beta_{1}}^{*}[S_{j}^{z}]
     A_{\alpha_{2},\beta_{2}}[S_{j}^{z}]\nonumber\\
=&
\begin{pmatrix}
  2& 0& 0& 2\\  0&-2& 0& 0\\
  0& 0&-2& 0\\  6& 0& 0& 2\\
\end{pmatrix}\nonumber.
\end{align}

An MPS representation is said to assume the canonical form~\cite{Vidal03,PerezGarcia07} 
when its transfer matrix satisfies the condition 
\begin{equation}
\begin{split}
 &\sum_{\alpha_{1},\alpha_{2}}\delta_{\alpha_{1},\alpha_{2}}
   T_{(\alpha_{1},\alpha_{2}),(\beta_{1},\beta_{2})}
   =\delta_{\beta_{1},\beta_{2}}\\
 &\sum_{\beta_{1},\beta_{2}}
   T_{(\alpha_{1},\alpha_{2}),(\beta_{1},\beta_{2})}
   \delta_{\beta_{1},\beta_{2}}
   =\delta_{\alpha_{1},\alpha_{2}}  \; .
\end{split}
 \label{eq:CanonicalCond}
\end{equation}
In order to discuss the SPT phase, it is convenient to work in 
the canonical form of the MPS. 
Since the MPS representation~(\ref{eq:MPS}) does not 
satisfy (\ref{eq:CanonicalCond}), 
we first render it canonical using a gauge transformation 
$A\to M^{-1}AM$ ($M$ is some matrix). 
It is obvious that this transformation does not change the MPS. 
Taking
\begin{equation}
 M=\begin{pmatrix} 3^{-1/4} & 0 \\  0 & 1 \end{pmatrix},\nonumber
\end{equation}
we obtain the following canonical form of the MPS: 
\begin{equation}
 |\Psi\rangle=\sum_{S_{j}^{z}=-3/2}^{3/2}
   \cdots \Lambda\Gamma[S_{j-1}^{z}]\Lambda\Gamma[S_{j}^{z}]
          \Lambda\Gamma[S_{j+1}^{z}]\Lambda\cdots
   \bigotimes_{j}|S_{j}^{z}\rangle,
 \label{eq:CanonicalMPS}
\end{equation}
where the matrices are given by
\begin{align}
 &\Gamma[ 3/2]=(1+\sqrt{3})^{-1/2}
   \begin{pmatrix} 0 & 0 \\ -\sqrt{2}\,3^{1/4} & 0 \end{pmatrix},\nonumber\\
 &\Gamma[ 1/2]=(1+\sqrt{3})^{-1/2}
   \begin{pmatrix}\sqrt{2}&0\\0&-\sqrt{2}\end{pmatrix},\nonumber\\
 &\Gamma[-1/2]=(1+\sqrt{3})^{-1/2}
   \begin{pmatrix} 0 & \sqrt{2}\,3^{1/4} \\ 0 & 0 \end{pmatrix},\nonumber\\
 &\Gamma[-3/2]=
   \begin{pmatrix} 0 & 0 \\  0 & 0 \end{pmatrix},\nonumber\\
 &\Lambda=\begin{pmatrix} 1/\sqrt{2} &0 \\ 0 & 1/\sqrt{2} \end{pmatrix}.
 \label{eq:LamGam}
\end{align}
The new MPS is related to the original one 
through the following 
similarity transformation: 
\begin{equation}
 M^{-1}A[S_{j}^{z}]M=(2+2\sqrt{3})^{1/2}
   \Lambda\Gamma[S_{j}^{z}] . \nonumber
\end{equation}
We can check the condition~(\ref{eq:CanonicalCond}) 
for the transfer matrix of the MPS (\ref{eq:CanonicalMPS}) 
\begin{align}
 T_{(\alpha_{1},\alpha_{2}),(\beta_{1},\beta_{2})}^{\rm can}\equiv &
   \sum_{S_{j}^{z}=-3/2}^{3/2}
     (\Lambda\Gamma)_{\alpha_{1},\beta_{1}}^{*}[S_{j}^{z}]
     (\Lambda\Gamma)_{\alpha_{2},\beta_{2}}[S_{j}^{z}]\nonumber\\
=&   \sum_{S_{j}^{z}=-3/2}^{3/2}
     (\Gamma\Lambda)_{\alpha_{1},\beta_{1}}^{*}[S_{j}^{z}]
     (\Gamma\Lambda)_{\alpha_{2},\beta_{2}}[S_{j}^{z}]\nonumber\\
=&\frac{1}{1+\sqrt{3}}
\begin{pmatrix}
  1& 0& 0& \sqrt{3}\\  0&-1& 0& 0\\
  0& 0&-1& 0\\  \sqrt{3}& 0& 0& 1\\
\end{pmatrix},\nonumber
\end{align}
and confirm that the MPS~(\ref{eq:CanonicalMPS}) is indeed in a canonical form. 

Now let us discuss the protecting symmetry. 
In contrast to the $S=1$ Haldane phase, time-reversal and $Z_{2}\times Z_{2}$ symmetries 
are explicitly broken due to the presence of an external magnetic field.  
Nevertheless, the system still retains inversion symmetry with respect to the center of a link ({\em link parity}) 
and, as is already suggested from the field-theory argument in Sec.\ref{sec:TemporalEdge}, 
this symmetry will do the job.  
According to the discussion in Refs.~\cite{Pollmann10,PerezGarcia08}, 
a projective representation $U_{\cal P}$ 
($\chi \times \chi$ unitary matrix) of link-parity $\mathcal{P}$ satisfies
\begin{equation}
 \Gamma=e^{i\theta_{\cal P}}
   U_{\cal P}^{\dagger}\Gamma^{\rm T}U_{\cal P}.
 \label{eq:LinkInvRep}
\end{equation}
Using (\ref{eq:LinkInvRep}) twice, we see that the fundamental property $\mathcal{P}^{2}=1$ 
of the link-center inversion implies 
\begin{align}
 &\Gamma=e^{2i\theta_{\cal P}}
   (U_{\cal P}^{*}U_{\cal P})^{\dagger}\Gamma(U_{\cal P}^{*}U_{\cal P})
 \nonumber\\
 \Leftrightarrow 
 &(U_{\cal P}^{*}U_{\cal P})\Gamma
   =e^{2i\theta_{\cal P}}\Gamma(U_{\cal P}^{*}U_{\cal P}).\nonumber
\end{align}
Multiplying $(\Gamma\Lambda)^{*}\Lambda$ from the left to this relation 
and taking the trace over the suffix $S^{z}$ in $\Gamma$, we obtain 
\begin{equation}
 (U_{\cal P}^{*}U_{\cal P})_{(\alpha_{1},\alpha_{2})}
   T_{(\alpha_{1},\alpha_{2}),(\beta_{1},\beta_{2})}^{\rm can}
   =e^{2i\theta_{\cal P}}
     (U_{\cal P}^{*}U_{\cal P})_{(\beta_{1},\beta_{2})}.\nonumber
\end{equation}
Here, note that $[U_{\cal P},\Lambda]=0$. 
Thus, $(U_{\cal P}^{*}U_{\cal P})$ is the left eigenvector 
of the transfer matrix $T^{\rm can}$. 
If we assume that unity is the unique largest eigenvalue of $T^{\rm can}$, 
then $e^{2i\theta_{\cal P}}=1$ and 
$U_{\cal P}^{*}U_{\cal P}=e^{i \phi_{\cal P}}E$ ($E$ is unit matrix). 
Using $U_{\cal P}=e^{i \phi_{\cal P}}U_{\cal P}^{\rm T}$ twice, 
$e^{2i \phi_{\cal P}}=1$. 
Thus, the phase $\phi_{\cal P}$ is quantized to 0 and $\pi$, 
which characterizes the SPT order protected by $\mathcal{P}$. 
In fact, $\phi_{\mathcal{P}}=0$ corresponds to a direct product (trivial) state. 
On the other hand, the state with $\phi_{\cal P}=\pi$, being characterized by a discrete integer, 
cannot be continuously deformed to the trivial one ($\phi_{\mathcal{P}}=0$) without a phase transition. 
From $U_{\cal P}=\pm U_{\cal P}^{\rm T}$, we can see that 
the unitary matrix $U_{\mathcal{P}}$ is either symmetric (trivial) or antisymmetric (topological). 

The structure of the entanglement spectrum also reflects the property of $U_{\cal P}$. 
Here we consider the topological case $\phi_{\cal P}=\pi$ 
(i.e., $U_{\cal P}=-U_{\cal P}^{\rm T}$). 
Since $U_{\cal P}$ and $\Lambda$ are commuting, the matrix $U_{\cal P}$ should be 
block diagonal according to subspaces labeled by 
the singular values $\lambda_{\alpha}$ (diagonal elements of $\Lambda$) 
which is equivalent to the entanglement eigenvalues. 
If we represent the dimension of the block $\alpha$ as $d_{\alpha}$, 
\begin{equation}
 \det(U_{\cal P})=\det(U_{\cal P}^{\rm T})
   =\det(-U_{\cal P})=(-1)^{d_{\alpha}}\det(U_{\cal P}).\nonumber
\end{equation}
Therefore, $d_{\alpha}$ should be even for any $\alpha$. 
This indicates that entanglement spectrum is 
twofold degenerate for SPT phases. 

For the model VBS state \eqref{eq:CanonicalMPS} for the $m=1/2$ plateau phase in the $S=3/2$ chain, 
we find that the following $\chi=2$ matrix 
\begin{equation}
 U_{\cal P}=i\sigma_{y}=
   \begin{pmatrix} 0 & 1 \\ -1 & 0 \end{pmatrix}
 \label{eq:MatrixLinkInv}
\end{equation}
satisfies Eq.~(\ref{eq:LinkInvRep}). 
As this is antisymmetric, we see that $\phi_{\cal P}$ is equal to $\pi$  
confirming that this plateau state is in the topological (Haldane) phase 
protected by link-center inversion symmetry $\mathcal{P}$.

It was assumed in the preceding arguments that the systems in question 
have U(1) rotational symmetry (at least) along the direction of 
the external field (i.e., $z$ axis). 
Therefore, the actual protecting symmetry is 
${\rm U(1)}\rtimes Z_{2}^{\rm P}$.  
In fact, a detailed comparison with the classification presented in Ref.~\cite{Chen12} 
suggests that our SPT plateaus may be embedded into the larger scheme 
based on the (cohomology) group $Z_{2}\times Z_{2}$.  
Within this classification scheme, there are three SPT phases as well as one trivial one, 
and the SPT plateau state that we have identified above corresponds to one of 
the three nontrivial phases. 
The detailed discussion on this ``larger picture" is provided in Appendix.

\section{Discussions and summary}
\label{sec:Summary}

Following a summary of what has been achieved thus far, 
we conclude by 
making several clarifying remarks on loose ends, 
and putting the work in context with related developments.

We began by deriving a semiclassical effective field theory 
which describes a magnetization plateau state. 
The action obtained was that of an XY model equipped with a topological term 
associated with space-time vortex events. Employing a path-integral 
formalism, we found that the latter term governs 
the structure of the ground state wave functional, 
and induces a topological distinction between plateau states 
with even and odd values of $S-m$.  
The source of this distinction may be tracked down to the fact that at the 
level of the original lattice model, only one in every two spatial links contributed 
to the total Berry phase, which in the continuum limit resulted  
in a crucial factor of 1/2. 
We further observed that an addition of a staggered magnetic field, which breaks 
the link-center inversion symmetry enjoyed by the original theory, destructs this 
topological distinction. A dual vortex field theory was derived which showed 
explicitly that with this perturbation the two 
previously distinct wave functionals 
could be smoothly deformed into each other without closing the energy gap.    
We were thus led to conclude that 
the $S-m=$ odd case is an SPT protected by link-center inversion symmetry, 
while the $S-m=$ even case is topologically trivial. 
An independent support for this expectation was provided through 
numerical calculations for $S=1/2$ and $S=1$ FFAF chains, 
as well as a rigorous treatment which incorporates an MPS 
representation of the magnetization plateau state. 

Readers may find it puzzling that we had started out with a lattice model 
which is identical to the one studied in Ref.~\cite{Tanaka09}, 
and yet we arrived at a different final expression for 
the topological term, which was crucial for what followed. 
The difference appeared from our  
adaption of the methods of Ref.~\cite{Sachdev02} originally devised for treating 
antiferromagnetic spin chains 
in the XY limit (and in the absence of a magnetic field) 
and retaining the surface contribution that inevitably arises once we employ this setup. 
We also took advantage of the fact that the bulk contribution to the Berry phase,  
being trivial in a magnetization plateau state, could safely be discarded. In short, 
the difference in outcome between the two treatments 
stems from the fact that in the present work 
we made use of a mathematical procedure which picks up the 
correct surface contributions to the spin Berry phases. 
Apart from the presence/absence of these surface terms, 
the two theories are equivalent. 

Our treatment of the ground state wave functional basically  
extends the prescriptions of  Ref.~\cite{Xu13} 
to (1+1) dimensions. 
The authors of  Ref.~\cite{Xu13}, in their investigation of SPT states 
arising out of a (3+1)-dimensional O(5) NL$\sigma$ model, needed to employ a limit in 
which the magnitude of one of the five components of their unit-length field was  
sent to zero. This intermediate step was necessary to derive a wave functional 
whose structure is governed by a topological obstruction at the (temporal) surface; 
in terms of the integer-valued winding number $N$ associated with this obstruction 
the ground state wave functional can take the nontrivial form $\Psi\propto (-1)^N$.  
Carrying out the corresponding procedure in (1+1) dimensions requires that we start with  
an O(3) NL$\sigma$ model and subsequently (i.e., after having expressed 
the wave functional in path-integral form) reduce the number of components to 
two. While this step of reducing the number of components appears 
somewhat artificial from a physical point of view, it arises naturally for our setup  
thanks to the presence of the external magnetic field which effectively kills the spin dynamics 
in the direction parallel to the applied field. In this regard the magnetization plateau phase 
in spin chains provides us with an ideal physical stage for studying  
possible SPT states along the ideas of Ref.~\cite{Xu13}.

\begin{figure}[t]
\includegraphics[width=0.45\textwidth]{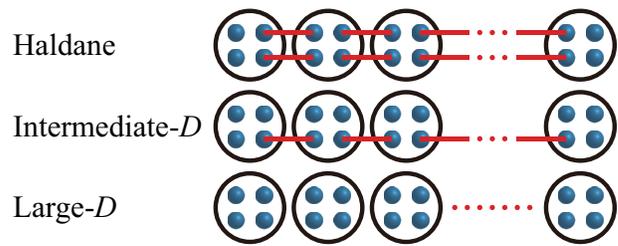}
\caption{(Color online) 
Schematic pictures of 
$S=2$ Haldane, intermediate-$D$, and large-$D$ phases.
The links connecting spin-1/2's (balls) 
represent singlet bonds.}
\label{fig:IntermediateD}
\end{figure}

It is also worth pointing out that our theory is also 
closely related to the physics of the so-called intermediate-$D$ phase,  
which can arise in quantum spin chains with $S\geq 2$ in the presence of a 
single ion anisotropy term (without an external magnetic field). 
This phase is expected~\cite{Oshikawa92} 
to lie in-between the Haldane gap phase 
(where, in the VBS picture, each one of the $2S$ auxiliary $S=1/2$ objects  
residing on a given site participates in the formation of a singlet bond with 
a neighboring site) and a trivial large-$D$ phase (where all 
$S=1/2$ objects are completely quenched out).  In this phase, only a 
limited number of $S=1/2$ spins per site are involved 
in the VBS-like structure, while the others are quenched out 
due to the anisotropy term.   
For the $S=2$ case, numerical studies have verified that 
a phase transition exists between 
the intermediate-$D$ and Haldane phases [Fig.~\ref{fig:IntermediateD}], 
while the Haldane and large-$D$ phases are connected 
within the phase diagram~\cite{Tonegawa11}. 
The intermediate-$D$ phase is analogous to the situation encountered 
in the preceding discussions in the sense that 
(i) the fluctuation of the spins occur predominantly within the $xy$ plane 
(assuming the anisotropic term is of the form $D(S_{j}^{z})^{2}$), and that 
(ii) the effective local spin moment is depleted 
owing to those $S=1/2$ objects 
which do not participate in the singlet bonds. We thus expect that the low-energy 
physics is again captured by the planar limit of the O(3) NL$\sigma$ model 
at the vacuum angle $\theta=2\pi(S-m)$ where $m$ is now the amount of 
local spin moment that is quenched.

Finally we mention that the field theoretical approach of  
Secs.~\ref{sec:EffectiveFieldTheory} and 
\ref{sec:TemporalEdge} also applies to the 
``Haldane insulators'' in bosonic systems~\cite{Fisher89,DallaTorre06,Batrouni13}. 
We start from a one-dimensional Bose-Hubbard model: 
\begin{equation}
 {\cal H}_{\rm BH}=-t\sum_{j}(b_{j}^{\dagger}b_{j+1}+{\rm H.c.})
   +\frac{U}{2}\sum_{j}n_{j}(n_{j}-1)-\mu\sum_{j}n_{j}\nonumber,
\end{equation}
where $b_{j}$, $b_{j}^{\dagger}$ are 
boson  
annihilation and creation 
operators, and $n_{j}\equiv b_{j}^{\dagger}b_{j}$ is 
a number operator. Switching to a 
coherent-state path integral language via the substitution  
$b_{j}(\tau)=\rho_{j}
^{1/2}e^{i\phi_{j}(\tau)}$ 
and writing $\rho_{j}=\rho_{0}+\delta\rho_{j}$ 
($\rho_{0}$ is the number of bosons per site), 
we obtain, upon integrating out $\delta\rho_{j}$ 
the effective Lagrangian~\cite{Herbut07} 
\begin{align}
 {\cal L}_{\rm BH}=&\frac{1}{2U}\sum_{j}\big(\partial_{\tau}\phi_{j}(\tau)\big)^{2}
   -t\rho_{0}\sum_{j}\cos\big[\phi_{j}(\tau)-\phi_{j+1}(\tau)\big]\nonumber\\
   &\quad+i\rho_{0}\sum_{j}\partial_{\tau}\phi_{j}(\tau).
 \label{eq:BHLagrangianEff}
\end{align} 
In the continuum limit 
the kinetic terms (the first two terms on the right-hand side) 
in (\ref{eq:BHLagrangianEff}) assume the form 
\begin{equation}
 {\cal L}_{\rm BH,kin}\sim\int dx
   \big[c_{\tau}\big(\partial_{\tau}\phi_{j}(\tau,x)\big)^{2}
       +c_{x   }\big(\partial_{x   }\phi_{j}(\tau,x)\big)^{2}\big],\nonumber
\end{equation}
where $c_{\tau}=1/(2Ua)$, $c_{x}=t\rho a/2$ ($a$ is the lattice constant). 
Moreover, the term $i\rho_{0}\sum_{j}\partial_{\tau}\phi_{j}(\tau)$ 
becomes identical with (\ref{eq:SiteBerryPhase}) 
upon the replacement $\rho_{0}\to S-m$. 
Therefore, the effective action for the Bose-Hubbard model 
takes the same form as (\ref{eq:PlateauEffAction}). 
Essentially repeating the arguments of  
Secs.~\ref{sec:EffectiveFieldTheory} and \ref{sec:TemporalEdge}, 
we are led to deduce that Haldane and trivial insulators 
in bosonic systems each  correspond to $\rho_{0}=$ odd and even, 
and that the addition of a staggered chemical potential 
(which is the counterpart of 
the staggered magnetization of our previous discussion) will destroy 
this topological distinction.

\acknowledgements
We thank Xiao-Gang Wen and Takahiro Morimoto 
for informative discussions, especially for 
helping us place this work in context with 
the general classification of SPT states. 
We also thank Muneto Nitta for discussions in the 
initial stage of this project. 
This work has been completed during the participation of S.T. 
in the long-term workshop ``Novel quantum states in condensed matter'' 
at Yukawa Institute for Theoretical Physics. 
This work is partially supported by Grants-in-Aid 
from the Japan Society for Promotion of Science, 
Grant No. (C) 23540461 (A.T.) and No. (C) 24540402 (K.T.). 
Numerical calculations were partially performed 
at the Supercomputer Center of the Institute for 
Solid State Physics, the University of Tokyo.

\appendix
\section{SPT phases protected by ${\rm U(1)}\rtimes Z_{2}^{\rm P}$ symmetry}
\label{sec:ProjectiveRep}

The SPT phases protected by the symmetry ${\rm U(1)}\rtimes Z_{2}^{\rm P}$ 
is classified by the cohomology group $Z_{2}\times Z_{2}$, 
as can be read off from the classification table~\cite{Chen12} 
for the mathematically equivalent entry ${\rm U(1)}\times Z_{2}^{\rm T}$.  
(Here $Z_{2}^{\rm T}$ and $Z_{2}^{\rm P}$ stand for time-reversal and link-parity 
symmetries, respectively.)
One of these two $Z_{2}$ groups corresponds to 
whether 
$U_{\cal P}^{*}U_{\cal P}=+1$ or $-1$ 
as is discussed in the main text, 
where $U_{\cal P}$ is a unitary matrix corresponding to the projective representation 
of the link-parity operation ${\cal P}$: 
\begin{equation}
\Gamma\xrightarrow{{\cal P}}\Gamma^{\text{T}}
=e^{-i\theta_{\cal P}}
   U_{\cal P} \Gamma^{\rm T}U_{\cal P}^{\dagger} \; .
 \label{eq:ParityInvRep}
\end{equation} 

Next, let us find the projective representation of the U(1) group, 
which consists of rotations around the $z$ axis with angle $\alpha$ 
($R_{z}^{(\alpha)}$) such as 
\begin{equation}
 R_{z}^{(\alpha)}\Gamma=e^{i\theta_{\alpha}}
   U_{z}^{(\alpha)\dagger}\Gamma U_{z}^{(\alpha)}.
 \label{eq:U1Rep}
\end{equation}
We consider the effect of exchanging the order in which the  
two operators $R_{z}^{(\alpha)}$ and ${\cal P}$ act.  
From (\ref{eq:ParityInvRep}) and (\ref{eq:U1Rep}), 
\begin{align}
 R_{z}^{(\alpha)}{\cal P}\Gamma
   =&e^{i(\theta_{\cal P}+\theta_{\alpha})}
     U_{z}^{(\alpha)\dagger}U_{\cal P}^{\dagger}\Gamma
     U_{\cal P}U_{z}^{(\alpha)}\nonumber\\
 {\cal P}R_{z}^{(\alpha)}\Gamma
   =&e^{i(\theta_{\cal P}+\theta_{\alpha})}
     U_{\cal P}^{\dagger}(U_{z}^{(\alpha)\dagger}\Gamma
     U_{z}^{(\alpha)})^{\rm T}U_{\cal P}.\nonumber\\
   =&e^{i(\theta_{\cal P}+\theta_{\alpha})}
     U_{\cal P}^{\dagger}U_{z}^{(\alpha){\rm T}}\Gamma
     U_{z}^{(\alpha)*}U_{\cal P}.\nonumber
\end{align}
The relation 
$R_{z}^{(\alpha)}{\cal P}\Gamma={\cal P}R_{z}^{(\alpha)}\Gamma$ 
leads to 
\begin{equation}
 \Gamma=U_{\cal P}U_{z}^{(\alpha)}U_{\cal P}^{\dagger}
   U_{z}^{(\alpha){\rm T}}\Gamma
   U_{z}^{(\alpha)*}U_{\cal P}
   U_{z}^{(\alpha)\dagger}U_{\cal P}^{\dagger}.\nonumber
\end{equation}
Hence, 
$U_{\cal P}U_{z}^{(\alpha)}U_{\cal P}^{\dagger}U_{z}^{(\alpha){\rm T}}$ 
should be the left eigenvector of the transfer matrix and is equal to
$e^{i\phi_{\alpha,{\cal P}}}E$ ($E$ is unit matrix). 
This implies that 
\begin{equation}
 U_{\cal P}U_{z}^{(\alpha)}
   =e^{i\phi_{\alpha,{\cal P}}}U_{z}^{(\alpha)*}U_{\cal P}.
 \label{eq:CommutRel}
\end{equation}
Since the action of U(1) is diagonal 
($U_{z}^{(\alpha){\rm T}}=U_{z}^{(\alpha)}$), 
$U_{z}^{(\alpha)*}=U_{z}^{(\alpha)\dagger}
=U_{z}^{(\alpha)\;-1}=U_{z}^{(-\alpha)}$. 
Using this relation and (\ref{eq:CommutRel}), 
we can also derive 
\begin{equation}
 U_{z}^{(\alpha)}U_{\cal P}
   =e^{-i\phi_{\alpha,{\cal P}}}U_{\cal P}U_{z}^{(\alpha)*}
 \label{eq:CommutRel2}
\end{equation}
by the replacement $\alpha\to-\alpha$. 
From (\ref{eq:CommutRel}) and (\ref{eq:CommutRel2}), 
$e^{i\phi_{\alpha,{\cal P}}}=\pm1$ is proved, i.e.,
$U_{\cal P}U_{z}^{(\alpha)}=\pm U_{z}^{(\alpha)*}U_{\cal P}$.
Therefore, the $Z_{2}\times Z_{2}$ classification 
of SPT phases protected by ${\rm U(1)}\rtimes Z_{2}^{\rm P}$ 
proceeds according to  
(i) $U_{\cal P}^{*}U_{\cal P}=+1$ or $-1$, and 
(ii) $U_{\cal P}U_{z}^{(\alpha)}=U_{z}^{(\alpha)*}U_{\cal P}$ 
or $U_{\cal P}U_{z}^{(\alpha)}=-U_{z}^{(\alpha)*}U_{\cal P}$. 

For the SPT plateau state discussed in 
Sec.~\ref{sec:MPS} (see Eq.~(\ref{eq:LamGam})), 
the U(1) rotation acts as 
\begin{align}
 &R_{z}^{(\alpha)}\Gamma[3/2]
   =e^{ 3i\phi_{\alpha}/2}\Gamma[3/2],\nonumber\\
 &R_{z}^{(\alpha)}\Gamma[1/2]
   =e^{  i\phi_{\alpha}/2}\Gamma[1/2],\nonumber\\
 &R_{z}^{(\alpha)}\Gamma[-1/2]
   =e^{ -i\phi_{\alpha}/2}\Gamma[-1/2],\nonumber\\
 &R_{z}^{(\alpha)}\Gamma[-3/2]
   =e^{-3i\phi_{\alpha}/2}\Gamma[-3/2].\nonumber
\end{align}
We can find that 
\begin{equation}
 U_{z}^{(\alpha)}=e^{i\alpha\sigma^{z}/2},\quad
 \theta_{\alpha}=\alpha/2
 \label{eq:MatrixU1}
\end{equation}
satisfies Eq.~(\ref{eq:U1Rep}). 
From (\ref{eq:MatrixLinkInv}) and (\ref{eq:MatrixU1}), 
we can confirm that 
$U_{\cal P}U_{z}^{(\alpha)}=U_{z}^{(\alpha)*}U_{\cal P}$ holds. 
Therefore, the SPT plateau belongs to the category with 
(i) $U_{\cal P}^{*}U_{\cal P}=-1$ and 
(ii) $U_{\cal P}U_{z}^{(\alpha)}=U_{z}^{(\alpha)*}U_{\cal P}$. 
The search for SPT phases with
$U_{\cal P}U_{z}^{(\alpha)}=-U_{z}^{(\alpha)*}U_{\cal P}$ 
is an interesting future problem.


\begin{thebibliography}{}

\bibitem{Wen04} 
X.-G. Wen, {\it Quantum Field Theory of Many-Body Systems} 
(Oxford University Press, Oxford, UK, 2004). 

\bibitem{Haldane83} 
F. D. M. Haldane, Phys. Lett. A {\bf 93}, 
\href{http://dx.doi.org/10.1016/0375-9601(83)90631-X}{464} (1983); 
\PRL{50}{1153}{1983}.

\bibitem{Affleck87} 
I. Affleck, T. Kennedy, E. H. Lieb, and H. Tasaki, \PRL{59}{799}{1987}; 
Commun. Math. Phys. {\bf 115}, \href{http://dx.doi.org/10.1007/BF01218021}{477} (1988).

\bibitem{Nightingale86}
M. P. Nightingale and H. W. J. Bl\"ote, \PRRC{B}{33}{659}{1986}.

\bibitem{White93}
S. R. White and D. A. Huse, \PR{B}{48}{3844}{1993}.

\bibitem{Polizzi98}
E. Polizzi, F. Mila, and E. S. S{\o}rensen, \PR{B}{58}{2407}{1998}.

\bibitem{Katsumata89}
K. Katsumata, H. Hori, T. Takeuchi, M. Date, A. Yamagishi, and J. P. Renard,
\PRL{63}{86}{1989}; 
Y. Ajiro, T. Goto, H. Kikuchi, T. Sakakibara, and T. Inami, 
{\it ibid.} {\bf 63}, \href{http://dx.doi.org/10.1103/PhysRevLett.63.1424}{1424} (1989).

\bibitem{Pollmann10} 
F. Pollmann, A. M. Turner, E. Berg, and M. Oshikawa, \PR{B}{81}{064439}{2010}; 
F. Pollmann, E. Berg, A. M. Turner, and M. Oshikawa, {\it ibid.} {\bf 85}, 
\href{http://dx.doi.org/10.1103/PhysRevB.85.075125}{075125} (2012). 

\bibitem{Chen11}
X. Chen, Z.-C. Gu, and X.-G. Wen, \PR{B}{84}{235128}{2011}.

\bibitem{Chen12}
X. Chen, Z.-C. Gu, Z.-X. Liu, and X.-G. Wen, Science {\bf 338}, 
\href{http://dx.doi.org/10.1126/science.1227224}{1604} (2012).

\bibitem{Levin12}
M. Levin and Z.-C. Gu, \PR{B}{86}{115109}{2012}.

\bibitem{Lu12}
Y.-M. Lu and A. Vishwanath, \PR{B}{86}{125119}{2012}.

\bibitem{Senthil13}
T. Senthil and M. Levin, \PRL{110}{046801}{2013}.

\bibitem{Vishwanath13}
A. Vishwanath and T. Senthil, \PR{X}{3}{011016}{2013}.

\bibitem{Ye13}
P. Ye and X.-G. Wen, \PR{B}{87}{195128}{2013}.

\bibitem{Lieb61}
E. H. Lieb, T. D. Schultz, and D. Mattis, Ann. Phys. (N. Y.) {\bf 16}, 
\href{http://dx.doi.org/10.1016/0003-4916(61)90115-4}{107} (1961).

\bibitem{Oshikawa97} 
M. Oshikawa, M. Yamanaka, and I. Affleck, \PRL{78}{1984}{1997}.

\bibitem{Totsuka97}
K. Totsuka, Phys. Lett. A {\bf 228}, 
\href{http://dx.doi.org/10.1016/S0375-9601(97)00087-X}{103} (1997).

\bibitem{Tanaka09}
A. Tanaka, K. Totsuka, and X. Hu, \PR{B}{79}{064412}{2009}.

\bibitem{Hu14}
H. Hu, C. Cheng, Z. Xu, H.-G. Luo, and S. Chen, 
\PR{B}{90}{035150}{2014}.

\bibitem{Takayoshi14}
S. Takayoshi, M. Sato, and T. Oka, \PR{B}{90}{214413}{2014}.

\bibitem{Sachdev11} S. Sachdev, {\it Quantum Phase Transitions} 
(Cambridge University Press, Cambridge, UK, 2011).

\bibitem{Fisher89}
M. P. A. Fisher, P. B. Weichman, G. Grinstein, and D. S. Fisher,
\PR{B}{40}{546}{1989}.

\bibitem{DallaTorre06}
E. G. Dalla Torre, E. Berg, and E. Altman, \PRL{97}{260401}{2006}.

\bibitem{Batrouni13}
G. G. Batrouni, R. T. Scalettar, V. G. Rousseau, and B. Gr\'emaud, 
\PRL{110}{265303}{2013}.

\bibitem{Endres11} 
M. Endres, M. Cheneau, T. Fukuhara, C. Weitenberg, P. Schau\ss, 
C. Gross, L. Mazza, M. C. Ba\~nuls, L. Pollet, I. Bloch, and S. Kuhr, 
Science {\bf 334}, 
\href{http://dx.doi.org/10.1126/science.1209284}{200} (2011). 

\bibitem{Xu13} 
C. Xu and T. Senthil, \PR{B}{87}{174412}{2013}.

\bibitem{Ng94} 
T.-K. Ng, \PR{B}{50}{555}{1994}.

\bibitem{Sachdev02}
S. Sachdev, Physica A {\bf 313}, 
\href{http://dx.doi.org/10.1016/S0378-4371(02)01040-3}{252} (2002).

\bibitem{Auerbach}
A. Auerbach, {\it Interacting Electrons and Quantum Magnetism} 
(Springer-Verlag, New York, 1994).

\bibitem{Fuji14}
Similar discussions are made also for symmetry-protected 
{\it trivial} phases in, e.g., 
Y. Fuji, F. Pollmann, and M. Oshikawa, 
\href{http://arxiv.org/abs/1409.8616}{arXiv:1409.8616}.

\bibitem{Affleck89}
I. Affleck, J. Phys.: Condens. Matter {\bf 1}, 
\href{http://dx.doi.org/10.1088/0953-8984/1/19/001}{3047} (1989).

\bibitem{Vidal07} 
G. Vidal, \PRL{98}{070201}{2007}.

\bibitem{Kitazawa00}
A. Kitazawa and K. Okamoto, \PR{B}{62}{940}{2000}.

\bibitem{Schulz86}
H. J. Schulz, \PR{B}{34}{6372}{1986}.

\bibitem{Chen03}
W. Chen, K. Hida, and B. C. Sanctuary, \PR{B}{67}{104401}{2003}.

\bibitem{Hida94}
K. Hida, \JPSJ{63}{2359}{1994}.

\bibitem{Li08}
H. Li and F. D. M. Haldane, \PRL{101}{010504}{2008}.

\bibitem{Nishioka09}
T. Nishioka, S. Ryu, and T. Takayanagi, J. Phys. A: Math. Theor. {\bf 42}, 
\href{http://dx.doi.org/10.1088/1751-8113/42/50/504008}{504008} (2009). 

\bibitem{footnote1}
Note that the lowest-energy state of Eq.~\eqref{eq:ParentHamMagnetizedVBS} 
is highly degenerate. 
Infinitesimally small magnetic field splits the degeneracy 
and selects the state Eq.~\eqref{eq:VBSPlateau} as the unique 
(up to edge states) ground state with bulk magnetization 1/2. 
The single-mode approximation predicts a finite gap 
to the excitation that changes magnetization by 1. 
Therefore, the magnetization curve jumps at $H=0$ to $m=1/2$ 
and has a finite plateau there. 

\bibitem{Vidal03}
G. Vidal, \PRL{91}{147902}{2003}. 

\bibitem{PerezGarcia07}
D. Perez-Garcia, F. Verstraete, M. M. Wolf, and J. I. Cirac,
Quantum Inf. Comput. {\bf 7}, 
\href{http://www.rintonpress.com/journals/qiconline.html#v7n56}{401} (2007).

\bibitem{PerezGarcia08}
D. Perez-Garcia, M. M. Wolf, M. Sanz, F. Verstraete, and J. I. Cirac, 
\PRL{100}{167202}{2008}.

\bibitem{Oshikawa92}
M. Oshikawa, J. Phys.: Condens. Matter {\bf 4}, 
\href{http://dx.doi.org/10.1088/0953-8984/4/36/019}{7469} (1992).

\bibitem{Tonegawa11}
T. Tonegawa, K. Okamoto, H. Nakano, T. Sakai, K. Nomura, and M. Kaburagi, 
\JPSJ{80}{043001}{2011}. 

\bibitem{Herbut07}
I. Herbut, {\it A Modern Approach to Critical Phenomena} 
(Cambridge University Press, Cambridge, UK, 2007).

\end{thebibliography}
\end{document}